\documentclass[reprint,aps,pra]{revtex4-1}

\usepackage{graphicx}
\usepackage{dcolumn}
\usepackage{bm}
\usepackage{amsmath}
\usepackage{amsthm}
\usepackage{amssymb}
\usepackage{float}
\usepackage{hyperref}
\usepackage{color}
\usepackage{epstopdf}
\usepackage{cleveref}
\usepackage[svgnames]{xcolor}
\hypersetup{hidelinks,colorlinks=true,allcolors=DarkBlue}

\begin{document}

\preprint{APS/123-QED}

\title{Minimization of errors in narrowband laser phase  \\ noise measurements based on reference measurement channels}
\author{A.B. Pnev$^{1}$}
\author{K.V. Stepanov$^{1}$}
\author{D.A. Dvoretskiy$^{1}$}
\author{A.A. Zhirnov$^{1}$}
\author{E.T. Nesterov$^{1}$}
\author{S.G. Sazonkin$^{1}$}
\author{\\ A.O. Chernutsky$^{1}$}
\author{D.A. Shelestov$^{1}$}
\author{A.K. Fedorov$^{1}$}
\author{C. Svelto$^{2}$}
\author{V.E. Karasik$^{1}$}
\affiliation
{
	\mbox{$^{1}$Bauman Moscow State Technical University, Moscow 105005, Russia}
	\mbox{$^{2}$Dipartimento di Elettronica e Informazione del Politecnico di Milano, Milano 20133, Italy}
}

\begin{abstract}
We propose a novel scheme for laser phase noise measurements with minimized sensitivity to external fluctuations including interferometer vibration, temperature instability, 
other low-frequency noise, and relative intensity noise. 
In order to minimize the effect of these external fluctuations, we employ simultaneous measurement of two spectrally separated channels in the scheme. 
We present an algorithm for selection of the desired signal to extract the phase noise. 
Experimental results demonstrate potential of the suggested scheme for a wide range of technological applications.

\end{abstract}

\date{\today}

\maketitle

\section{Introduction}

Intrinsic phase noise in a laser oscillator is caused by spontaneous emission of photons propagating collinearly to stimulated emission radiation. 
This type of noise contributes to the output laser spectrum. 
As a result, one can observe broadening of the spectrum and appearance of side peaks around the main optical frequency. 
Phase noise can be described as a continuous frequency drift in combination with spontaneous phase jumps~\cite{Henry}.
Quantum noise, such as spontaneous emission of gain media, is the fundamental reason of the existence of phase noise. 
In addition to its basic quantum origin, the laser phase noise can be also generated by technical noise contributions, 
{\it e.g.} due to cavity vibrations and temperature fluctuations. 
In many cases the phase noise is connected to the relative intensity noise (RIN) of the laser~\cite{Derickson}.

There are two important tasks in phase noise measurements. 
The first is to eliminate external influences. 
The second one is providing sufficient sensitivity of registration system for proper signal detection. 
Precise phase noise measurements are especially important for lasers designed as a light source for high speed binary phase-shift keying (BPSK) 
and quadrature phase-shift keying (QPSK) optical networks, 
LIght Detection And Ranging (LIDAR), phase-sensitive optical time-domain reflectometry, and fiber optic hydrophones.

Phase noise of a laser source plays a crucial role for efficient operating of these systems. 
In the present work, we suggest a novel approach for phase noise measurements providing both theoretical and experiment analysis. 
The suggested approach is based on the absence of phase modulator and feedback loop.
These parts complicate common schemes for phase noise measurements~\cite{Llopis,Augst,Zhu,Williams,Meng,Ivo,Kostromitin}, and they are additional source of errors. 
We demonstrate that the proposed scheme allows phase noise measurement in the range of $0.1\dots10$ Hz from the carrier optical frequency, 
whereas precise noise measurement in this range is challenging~\cite{Stolpner}. 
This range is of significant interest for seismic sensors in areas of permafrost and hydrophones~\cite{Costley}. 
The proposed scheme has a potential for application in phase noise measurements of lasers with uncompensated relaxation oscillations in a frequency range less than 1 MHz~\cite{Zhirnov} 
due to taking into account RIN for its further compensation.

\begin{figure}
\includegraphics[width=1\linewidth]{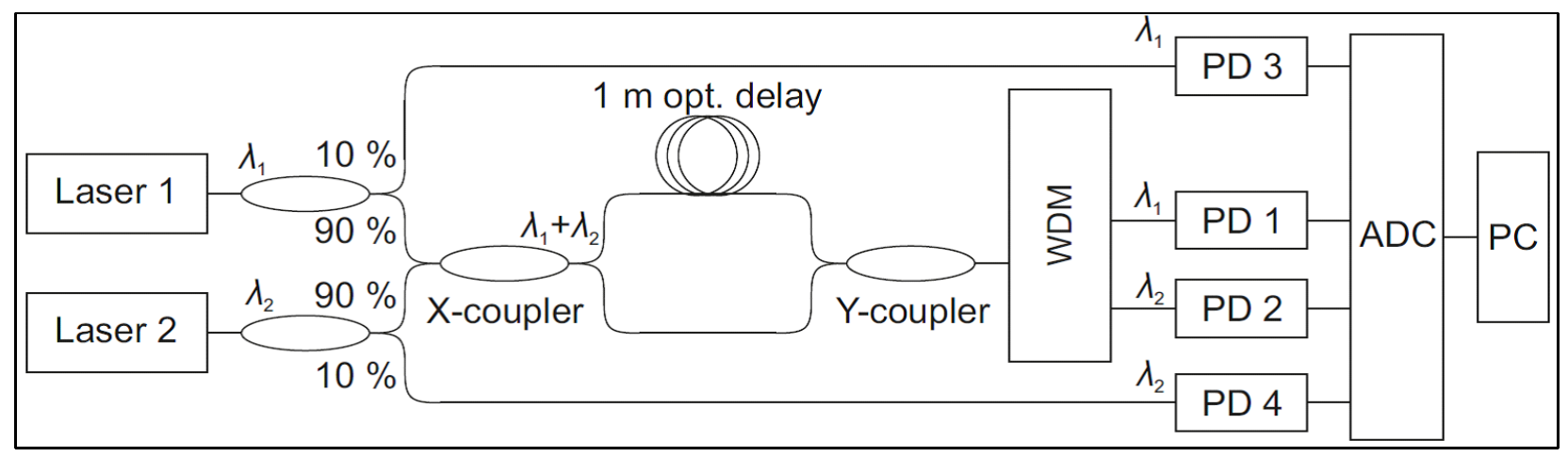}
\vskip -4mm
\caption
{
Scheme of the experimental setup.
} 
\label{fig:setup}
\end{figure}

\section{Theoretical analysis}

The scheme of the proposed measurement setup is given in Fig.~\ref{fig:setup}. 
We describe its basic working principles on the example of work of Laser 1. Light from Laser 1 goes to a coupler. A small part (about 10\%) is forwarded to the photodiode PD3, 
the other part (about 90\%) goes to the unbalanced Mach-Zehnder interferometer (MZI), created by X-coupler, Y-coupler, 
and two fiber arms (one of the arms has longer optical delay of 1 m)~\cite{Derickson}.

The radiation from output of the Y-coupler goes to the wave-division multiplexer (WDM) and is then forwarded to the photodiode PD 1 in accordance to its wavelength. 
For Laser 2 the procedure is the same and light finally comes to photodiodes PD 4 and PD 2.
Signal from photodiodes converts to the digital form by the ADC and then goes to the processing in a PC.

\begin{figure}
\includegraphics[width=1\linewidth]{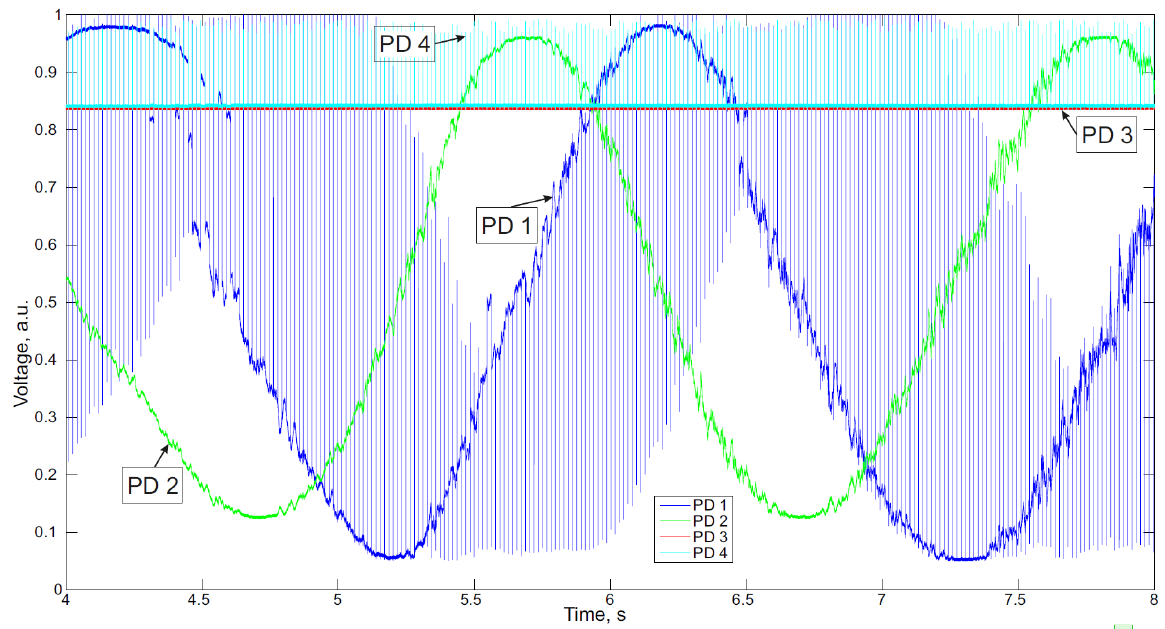}
\vskip -4mm
\caption
{
Time evolution of experimental data from receivers PD 1 --- PD 4.
} 
\label{fig:evolution}
\end{figure}

Transformations of signals in the setup can be described via coupling coefficients and phase delays in the interferometer. 
If we take $L_1$ as the intensity of Laser 1, then $0.1L_1$ is the intensity level on PD 3, and $0.9L_1$ is the intensity at the MZI input. 
The intensity is divided in two arms in equal proportions: $P_1=P_2=0.45L_1$.
The 1 m optical delay is too small to create sufficient intensity difference. 
At the output we receive the constant part $P_1+P_2$ and the variable part with maximum amplitude $2\sqrt{({P_1}\times{P_2})}$. 
Last term is defined by phase difference created by optical delay and calculated as $\exp{\left[-2\pi{i}(l+\delta{l})/(\lambda+\delta\lambda)\right]}$,
where $l=1$ m is optical delay, $\Delta{l}$ is its deviations, $\lambda=1550$ nm is the laser wavelength, and $\Delta{\lambda}$ is its fluctuations. 
Finally, we can present signal transformations in the MZI by the following transfer function~\cite{Hui}:
\begin{equation}\label{eq:1}
\begin{split}
	\!F_{\rm MZI}(\Delta,\nu)&=P_1+P_2+2\sqrt{({P_1}\times{P_2})}\times\\
	&\times\cos\left(\frac{2\pi{\nu}}{c}\Delta\right)\sim\cos^2\left(\frac{\pi{\nu}}{c}\Delta\right),
\end{split}
\end{equation}
where $\Delta$ is the optical length between interferometer arms, $c$ is the speed of light, and $\nu$ is the frequency of the laser.
Received signal through WDM goes to photodiodes with sensitivity at 1550 nm equal to $0.9$ A/W, so we get ${\rm PD}_1=0.9F_{\rm MZI}$,  
and ${\rm PD}_3=0.09L_1$. 
On the next step after ADC we receive discrete quantized signal:
\begin{equation}
\begin{split}
	PD_{1d}(t_i)=\left\lfloor\frac{PD_1(t)\times{\rm comb}(t/T_d)}{U_{\rm ref}}\right\rfloor, \\
	PD_{3d}(t_i)=\left\lfloor\frac{PD_3(t)\times{\rm comb}(t/T_d)}{U_{\rm ref}}\right\rfloor.
\end{split}
\end{equation}
Here $t$ is the continuous measurement time, $t_i$ is the discrete time of $i$th step, $T_d$ is the ADC discretization time, 
and $U_{\rm ref}$ is the ADC reference level. 
For the second signal transformations are described in the same way for PD 2 and PD 4.

In the proposed setup, phase noise of each spectrally separated laser source can be measured in parallel. 
Then they should be different in optical frequencies at least on the value about 100 GHz of optical frequencies due to difference in adjacent channels in WDM for optical networks. 
Our scheme also requires sufficient level of stability of the lasers and absence of the exchange in optical spectral channels (e.g. channels of ITU grid).
Phase noise is calculated from obtained data by the following algorithm:
\begin{enumerate}
	
	\item Data acquisition from photodiodes.
	
	Array of MZI intensity values for Laser 1: $M1$.
		
	Array of MZI intensity values for Laser 2: $M2$.
	
	Array of intensity values for Laser 1 directly: $M3$.
		
	Array of intensity values for Laser 2 directly: $M4$.

	\item Obtaining $F1$ and $F2$ by calculating the Fourier transforms of $M1$ and $M2$.

	\item $F1$ and F2 comparison in low frequency range (from $0.1$ to $2$ Hz) and definition of their common part LF. 
	This part is caused mainly by MZI instability and much less by laser intensity noises.
	
	\item Filtration of defined in the previous paragraph low frequency part $LF$ (from $0.1$ to $2$ Hz) in the arrays $F1$ and $F2$ via a low frequency filter.
	
	\item Calculation of filtered time-domain signals $M1f$ and $M2f$ via the inverse Fourier transform of $F1$ and $F2$.
	
	\item *Next points describe transformation of $M1f$ and $M3$ into phase noise of Laser 1. 
	Phase noise of Laser 2 can be calculated in the same way.

	\item Compensation of laser intensity instability:

		\begin{enumerate}
		
		\item determination of the proportionality factor $K1=\langle{M1f}\rangle/\langle{M3}\rangle$, 
		which describes intensity ratio of Laser 1 power in PD 1 and PD 3 arms of the setup;
		
		\item subtraction of the constant component ($P_1$ and $P_2=P_1$ in Eq.~(\ref{eq:1})): 
		
		$$
			M1k={M1f}-{M3}\times{K1};
		$$
	
		\item calculation of interference term of Eq. (1) without its intensity-dependent amplitude 
		$$
			2\sqrt{({P_1}\times{P_2})}:M1h=M1k/({M3}\times{K1}).
		$$
		
		\end{enumerate}
		
	\item Retrieval of phase fluctuations:
		\begin{equation}
			Ph1=\arcsin(M1h).
		\end{equation}
		
	\item Calculation of phase fluctuations spectrum taking into account frequency resolution ($BW=0.1$ Hz defined by measurement time) and constant phase lag from $l=1$ m optical delay:
	$\phi=2\pi{L}/\lambda$, where $\lambda$ is the lasers wavelength, $L=l+\delta{l}\approx{l}$ since $l\gg{\delta{l}}$, and $\delta{l}$ includes external influences and measurement error:
		\begin{equation}
			P_f=\frac{\left|\mathcal{F}[Ph1]\right|}{BW\phi^2},
		\end{equation}
		where $\mathcal{F}[Ph1]$ is the Fourier transform operator.
	\item Phase noise is defined as a square root from power spectrum:
		\begin{equation}
			\Phi_N=\sqrt{P_f}.
		\end{equation}
\end{enumerate}

The proposed method allows to measure phase noise in a frequency noise down to $0.1$ Hz and with a frequency resolution $RBW=0.1$ Hz.

\section{Error analysis} 

\begin{figure}
\includegraphics[width=0.9\linewidth]{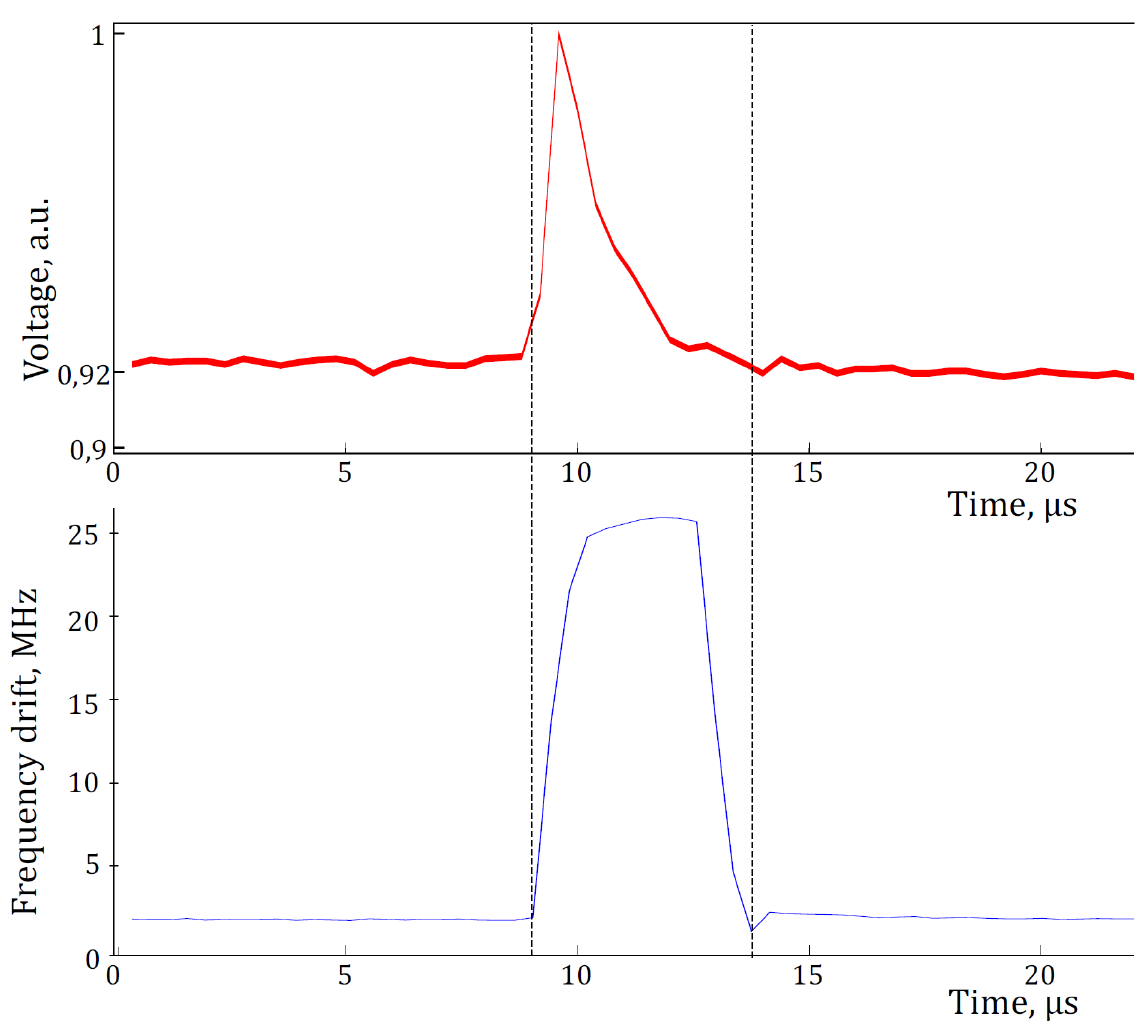}
\vskip -4mm
\caption
{
Graph showing influence of laser intensity instability on the laser frequency.
} 
\label{fig:graph}
\end{figure}

We calculate the impact of the following factors: 
temperature drift, optical path difference, and central wavelength instability. 
To this goal, we use lasers with central wavelengths $\lambda_1=1550.12$ nm ($f_1=193.40$ THz) and $\lambda_2=1549.32$ nm ($f_2=193.50$ THz), 
which correspond to channels 34 and 35 of ITU Grid: C-Band.

The temperature instability causes drifts in the optical path difference respectively, which in turn change the interferometer arms difference:
\begin{equation}
	\Delta=\left(\alpha_{\Lambda}+\alpha_{n}\right)l\Delta{t},
\end{equation}
where $\alpha_{\Lambda}$ is the thermal expansion coefficient ($\alpha_{\Lambda}=0.55\times10^{-6}$ fused silica),
$\alpha_{n}$ is the thermo-optic coefficient ($\alpha_{n}=8.6\times10^{-6}$ for $\lambda=1550$ nm) \cite{Meltz},
$l$ is the length of expansion area, in our case $l$ is the interferometer arms difference ($l=1$ m). 

The temperature instability of $0.1^{\circ}$ causes phase drifts:
\begin{equation}
\begin{split}
	\Delta\phi_1=\frac{2\pi}{\lambda_1}\Delta=1.2902\pi,\\
	\Delta\phi_2=\frac{2\pi}{\lambda_2}\Delta=1.2909\pi.
\end{split}
\end{equation}

Several receivers carry out registration of signals. 
That is why the emission path of Laser 1 to PD 1 and PD 3 (Laser 2 emission path to PD 2 and PD 4) should be the same in order to minimize calculation errors. 
Otherwise, the compensation of the amplitude noise is incorrect. 
Precision of this path can be 1 cm using known length patch cord.

We conduct an analysis of the wavelength drift impact on registered signal. 
We assume that there is no change in interferometer arms difference. 
Then the transmission function (\ref{eq:1}) makes a complete oscillation under the condition:
\begin{equation}
	\frac{2\pi}{\lambda}\Delta-\frac{2\pi}{\lambda+\Delta\lambda}\Delta=2\pi.
\end{equation}
Thus, for wavelengths $\lambda_1{=}1550.12$ nm and $\lambda_2{=}1549.32$ nm the complete oscillation of the transmission function takes places at wavelength drift values 
$\Delta\lambda_1=2.4028$ pm and $\Delta\lambda_2= 2.4003$ pm, respectively.
		
\section{Experimental investigation} 

To illustrate our results we implemented on experimental study of the suggested scheme. 
We use the following lasers: Redfern Integrated Optics-OptaSense Inc. (Laser 1) and DenseLight Semiconductors (Laser 2). 
The interference signals received at photodetector outputs during the experiment are shown in Fig.~\ref{fig:evolution}.

One interesting feature of Laser 2 (DenseLight) is a random intensity jumps. 
That is why the curve of Laser 2 has highly noisy data. 
The effect of the intensity instability of Laser 2 on the frequency is shown in Fig.~\ref{fig:graph}. 
One can see that frequency instability is longer then intensity instability. 
Consideration of theoretical base for this phenomenon is out of the scope of this article. 
Laser 1 (Redfern Integrated Optics) does not have short intensity jumps that is why its experimental curve is smooth. 
It changes only due to the external factors (vibration, temperature changes, and etc.) and measured phase noise.

\begin{figure}
\includegraphics[width=0.9\linewidth]{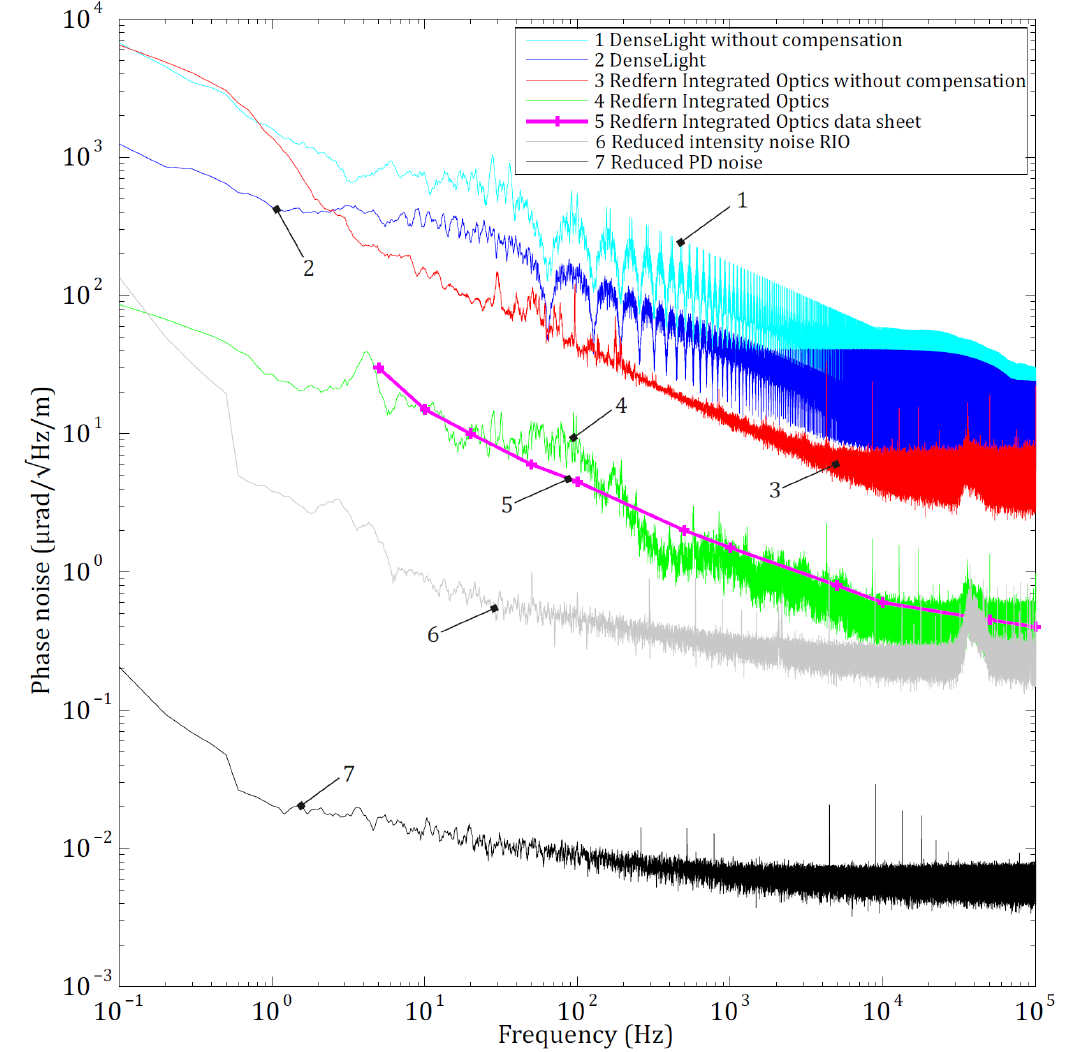}
\vskip -2mm
\caption
{
Phase noise plots: 
curves 1 and 3 are without proposed method usage,
curves 2 and 4 are with proposed method usage,
curve 5 is the Laser 1 specification data,
curve 6 is the reduced intensity noise,
curve 7 is the reduced photoreceiver noise.
} 
\label{fig:noise}
\end{figure}

Experimental curves of measured phase noise spectra using the proposed method for two lasers are presented in Fig.~\ref{fig:noise}. 
For comparison also we show the results without low frequency compensation (curves 1 and 3 in Fig.~\ref{fig:noise}). 
Spectral densities without compensation (curves 2 and 4 in Fig.~\ref{fig:noise}) are much higher than ones thou the ones obtained after proper compensation. 
Thus, we can conclude that due to high MZI sensitivity to external perturbations spectral compensation is extremely necessary. 
We also show a phase noise curve from the Redfern Integrated Optics laser (specification is given in Ref.~\cite{Rio}) 
(curve 5 in Fig.~\ref{fig:noise}) obtained by the setup described in Ref.~\cite{Stolpner}. 
It is clear that results achieved with our setup are in good agreement with these data. 
We show the different system fluctuations reduced to phase noise: intensity noise of Redfern Integrated Optics laser (curve 6 in in Fig.~\ref{fig:noise}) 
and photoreceiver background noise (curve 7 in Fig.~\ref{fig:noise}). 
Analysis of curves 6 and 4 also explains a broad peak in $30\dots50$ kHz range, since it is the laser intensity noise not connected with laser wavelength. 
Analysis of curves 7 and 4 allows one to explain narrow peaks in the high frequency region. 
They are noises of the photoreceiver, which is not connected to the laser. 
Curves 6 and 7 allow understanding the noise floor of the measurement setup. 
One can see that photoreceiver noise gives a very little contribution. 
The laser intensity instability remains the most important error source after MZI fluctuation compensation.

\section{Conclusion}

A novel laser phase noise measurement scheme was described. 
We showed that the proposed scheme allows avoiding use of phase modulators and provides minimizing of errors due to vibrations and temperature instability. 
The proposed algorithm to determine the phase noise allows measuring phase noise in the low-frequency range up to sub-Hz was presented and experimentally tested.
The result achieved by compensating the same components from two spectral demultiplexing narrowband laser sources were presented. 
Experimental data from the proposed scheme are in good agreement with specification. 
These merits make our scheme a very promising solution.

\section{Acknowledgement}

We thank S.V.\,Tikhomirov for useful comments.
This work was supported under the project 14.577.21.0224 (Unique identifier RFMEFI57716X0224) from the Ministry of Education and Science of the Russian Federation.

\end{document}